\documentclass[12pt]{article}
\usepackage{amsmath}
\usepackage{graphicx}
\begin{document}
\baselineskip=18 pt
\begin{center}
{\large{\bf Effects of Kaluza-Klein theory and potential on a generalized Klein-Gordon oscillator in the cosmic string space-time }}
\end{center}

\vspace{.5cm}

\begin{center}
{\bf Faizuddin Ahmed}\footnote{faizuddinahmed15@gmail.com ; faiz4U.enter@rediffmail.com}\\ 
{\bf Ajmal College of Arts and Science, Dhubri-783324, Assam, India}
\end{center}

\vspace{.5cm}

\begin{abstract}

In this paper, we solve a generalized Klein-Gordon oscillator in the cosmic string space-time with a scalar potential of Cornell-type within the Kaluza-Klein theory and obtain the relativistic energy eigenvalues and eigenfunctions. We extend this analysis by replacing the Cornell-type with Coulomb-type potential in the magnetic cosmic string space-time and analyze a relativistic analogue of the Aharonov-Bohm effect for bound states.

\end{abstract}

\vspace{.3cm}

{\bf keywords:} Klein-Gordon oscillator, topological defects, Kaluza-Klein theory, special functions, Aharonov-Bohm effect.

\vspace{0.3cm}

{\bf PACS Number(s):} 03.65.Pm, 03.65.Ge, 04.50.Cd, 04.50.+h, 11.10.Kk

\section{Introduction}

A unified formulation of Einstein's theory of gravitation and theory of electromagnetism in four-dimensional space-time was first proposed by Kaluza \cite{bb12} by assuming a pure gravitational theory in five-dimensional space-time. The so called cylinder condition was later explained by Klein when the extra dimension was compactified on a circle $S^1$ with a microscopic radius \cite{bb13}, where the spatial dimension becomes five-dimensional. The idea behind introducing additional space-time dimensions has found application in quantum field theory, for instance, in string theory \cite{bb26}. There are studies on Kaluza-Klein theory with torsion \cite{bb31,bb32}, in the Grassmannian context \cite{ss1,ss2,ss3}, in K\"{a}hler fields \cite{ss4}, in the presence of fermions \cite{bb33,bb34,bb35}, and in the Lorentz-symmetry violation (LSV) \cite{bb36, bb37,bb38}. Also, there are investigations in space-times with topological defect  in the context of Kaluza-Klein theory, for example, the magnetic cosmic string \cite{bb14} (see also, \cite{ss5}), and magnetic chiral cosmic string \cite{bb28} in five-dimensions.

Aharonov-Bohm effect \cite{bb39,bb40,bb50} is a quantum mechanical phenomena that has been investigated in several branches of physics, such as in, graphene \cite{RJ}, Newtonian theory \cite{MAA}, bound states of massive fermions \cite{VRK}, scattering of dislocated wave-fronts \cite{CC}, torsion effect on a relativistic position-dependent mass system \cite{ff3,AHEP}, non-minimal Lorentz-violating coupling \cite{HB}. In addition, Aharonov-Bohm effect has been investigated in the context of Kaluza-Klein theory by several authors \cite{bb28,bb15,bb16,aa6,EVBL,EVBL2,EPJC}, and the geometric quantum phase in graphene \cite{KB}. It is well-known in condensed matter \cite{NB,WCT,LD,MBu,ACB} and in the relativistic quantum systems \cite{Bakke,Bakke2} that when there exist dependence of the energy eigenvalues on geometric quantum phase \cite{bb50}, then, persistent current arise in the systems. The studies of persistent currents have explored systems that deal with the Berry phase \cite{DL,DL2}, the Aharonov-Anandan quantum phase \cite{XCG,TZQ}, and the Aharonov-Casher geometric quantum phase \cite{AVB,SO,HM2,HM3}. Investigation of magnetization and persistent currents of mass-less Dirac Fermions confined in a quantum dot in a graphene layer with topological defects was studied in \cite{cc17}.

Klein-Gordon oscillator theory \cite{bb1,bb2} was inspired by the Dirac oscillator \cite{bb3}. This oscillator field is used to study the spectral distribution of energy eigenvalues and eigenfunctions in $1-d$ version of Minkowski space-times \cite{bb4}. Klein-Gordon oscillator was studied by several authors, such as, in the cosmic string space-time with  an external fields \cite{bb5}, with Coulomb-type potential by two ways : (i) modifying the mass term $m \rightarrow m+S(r)$ \cite{bb6}, and (ii) via the minimal coupling \cite{bb7} in addition to a linear scalar potential, in the background space-time generated by a cosmic string \cite{bb8}, in the G\"{o}del-type space-times under the influence of gravitational fields produced by topological defects \cite{bb9}, in the Som-Raychaudhuri space-time with a disclination parameter \cite{ff2}, in non-commutative (NC) phase space \cite{bb10}, in $(1+2)$-dimensional G\"{u}rses space-time background \cite{ff4}, and in $(1+2)$-dimensional G\"{u}rses space-time background subject to a Coulomb-type potential \cite{ff5}. The relativistic quantum effects on oscillator field with a linear confining potential was investigated in \cite{ff6}. 

We consider a generalization of the oscillator as described in Refs. \cite{EPJC,ff5} for the Klein-Gordon. This generalization is introduced through a generalized momentum operator where the radial coordinate $r$ is replaced by a general function $f (r)$. To author best knowledge, such a new coupling was first introduced by K. Bakke {\it et al.} in Ref. \cite{Bakke} and led to a generalization of the Tan-Inkson model of a two-dimensional quantum ring for systems whose energy levels depend on the coupling's control parameters. Based on this, a generalized Dirac oscillator in the cosmic string space-time was studied by F. Deng {\it et al.} in Ref. \cite{cc9} where the four-momentum $p_{\mu}$ is replaced with its alternative $p_{\mu}+m\,\omega\,\beta\,f_{\mu} ( x_{\mu} )$. In the literature, $f_{\mu} (x_{\mu})$ has chosen similar to potentials encountered in quantum mechanics (Cornell-type, exponential-type, singular, Morse-type, Yukawa-like etc.). A generalized Dirac oscillator in $(2+1)$-dimensional world was studied in \cite{cc10}. Very recently, the generalized K-G oscillator in the cosmic string space-time in \cite{FD}, and non-inertial effects on a generalized DKP oscillator in the cosmic string space-time in \cite{SZ} was studied.

The relativistic quantum dynamics of a scalar particle of mass $m$ with a scalar potential $S (r)$ \cite{bb41,WG} is described by the following Klein-Gordon equation:
\begin{equation}
\left [\frac{1}{\sqrt{-g}}\,\partial_{\mu} (\sqrt{-g}\,g^{\mu\nu}\,\partial_{\nu})-(m + S)^2 \right]\,\Psi=0,
\label{1}
\end{equation}
with $g$ is the determinant of metric tensor with $g^{\mu\nu}$ its inverse. To couple Klein-Gordon field with oscillator \cite{bb1,bb2}, following change in the momentum operator is considered as in \cite{dd2,bb5}:
\begin{equation}
\vec{p}\rightarrow \vec{p}+i\,m\,\omega\,\vec{r},
\label{2}
\end{equation}
where $\omega$ is the oscillatory frequency of the particle and $\vec{r}=r\,\hat{r}$ where, $r$ being distance from the particle to the string. To generalized the Klein-Gordon oscillator, we adopted the idea considered in Refs. \cite{EPJC,ff5,cc9,FD,SZ} by replacing $r \rightarrow f (r)$ as
\begin{equation}
X_{\mu}=(0, f(r), 0, 0, 0).
\label{3}
\end{equation}
So we can write $\vec{p}\rightarrow \vec{p}+i\,m\,\omega\,f (r)\,\hat{r}$, and we have, $p^2 \rightarrow  (\vec{p}+i\,m\,\omega\, f (r)\,\hat{r})(\vec{p}-i\,m\,\omega\,f (r)\,\hat{r})$. Therefore, the generalized Klein-Gordon oscillator equation:
\begin{equation}
\left[\frac{1}{\sqrt{-g}}\,(\partial_{\mu}+m\,\omega\,X_{\mu})\{\sqrt{-g}\,g^{\mu\nu}\,(\partial_{\nu}-m\,\omega\,X_{\nu})\}-(m+ S)^2 \right]\,\Psi=0,
\label{4}
\end{equation}
where $X_{\mu}$ is given by Eq. (\ref{3}). 

Various potentials have been used to investigate the bound state solutions to the relativistic wave-equations. Among them, much attention has given on Coulomb-type potential. This kind of potential has widely used to study various physical phenomena, such as in, the propagation of gravitational waves \cite{gg}, the confinement of quark models \cite{gg1}, molecular models \cite{gg2}, position-dependent mass systems \cite{gg3,gg4,gg5}, and relativistic quantum mechanics \cite{bb7,bb8,bb6}. The Coulomb-type potential is given by
\begin{equation}
S(r)=\frac{\eta_{c}}{r}.
\label{5}
\end{equation}
where $\eta_{c}$ is Coulombic confining parameter.

Another potential that we are interest here is the Cornell-type potential. The Cornell potential, which consists of a linear potential plus a Coulomb potential, is a particular case of the quark-antiquark interaction, one more harmonic type term \cite{gg6}. The Coulomb potential is responsible by the interaction at small distances and the linear potential leads to the confinement. Recently, the Cornell potential has been studied in the ground state of three quarks \cite{CA}. However, this type of potential is worked on spherical symmetry; in cylindrical symmetry, which is our case, this type of potential is known as Cornell-type potential \cite{bb9}. This type of interaction has been studied in \cite{bb9,ff6,bb41,RLLV}. Given this, let us consider this type of potential 
\begin{equation}
S(r)=\eta_{L}\,r+\frac{\eta_c}{r},
\label{6}
\end{equation}
where $\eta_{L}, \eta_{c}$ are the confining potential parameters.

The aim of the present work is to analyze a relativistic analogue of the Aharonov-Bohm effect for bound states \cite{bb39,bb40,bb50} for a relativistic scalar particle with potential in the context of Kaluza-Klein theory. First, we study a relativistic scalar particle by solving the generalized Klein-Gordon oscillator with a Cornell-type potential in the five-dimensional cosmic string space-time. Secondly, by using the Kaluza-Klein theory \cite{bb12,bb13,bb26} a magnetic flux through the line-element of the cosmic string space-time is introduced, and thus write the generalized Klein-Gordon oscillator in the five-dimensional space-time. In the later case, a Coulomb-type potential by modifying the mass term $m \rightarrow m + S(r)$ is introduced which was not study earlier. Then, we show that the relativistic bound states solutions can be achieved, where the relativistic energy eigenvalues depend on the geometric quantum phase \cite{bb50}. Due to this dependence of the relativistic energy eigenvalue on the geometric quantum phase, we calculate the persistent currents \cite{NB,WCT} that arise in the relativistic system.

This paper comprises as follow : In {\it section 2}, we study a generalized Klein-Gordon oscillator in the cosmic string background within the Kaluza-Klein theory with a Cornell-type scalar potential; in {\it section 3}, a generalized Klein-Gordon oscillator in the magnetic cosmic string in the Kaluza-Klein theory subject to a Coulomb-type scalar potential and obtain the energy eigenvalues and eigenfunctions; and the conclusion one in {\it section 4}.

\section{Generalized Klein-Gordon oscillator in cosmic string space-time with a Cornell-type potential in Kaluza-Klein theory}

The purpose of this section is to study the Klein-Gordon equation in cosmic string space-time with the use of Kaluza-Klein theory with interactions. The first study of the topological defects within the Kaluza-Klein theory was carried out in \cite{bb14}. The metric corresponding to this geometry can be written as,
\begin{equation}
ds^2=-dt^2+dr^2+\alpha^2\,r^2\,d\phi^2+dz^2+dx^2,
\label{7}
\end{equation}
where $t$ is the time coordinate, $x$ is the coordinate associated with the fifth additional dimension and $(r, \phi, z)$ are cylindrical coordinates. These coordinates assume the ranges $-\infty < (t, z) < \infty$, $0 \leq r < \infty$, $0 \leq \phi \leq 2\,\pi$, $0 < x < 2\,\pi\,a$, where $a$ is the radius of the compact dimension $x$. The $\alpha$ parameter characterizing the cosmic string, and in terms of mass density $\mu$ given by $\alpha=1-4\,\mu$ \cite{bb21}. The cosmology and gravitation imposes limits to the range of the $\alpha$ parameter which is restricted to $\alpha <1$ \cite{bb21}.

By considering the line element (\ref{7}) into the Eq. (\ref{4}), we obtain the following differential equation :
\begin{eqnarray}
&&[-\frac{\partial^2}{\partial t^2}+\frac{1}{r}\,\left(\frac{\partial}{\partial r} + m\,\omega\,f (r) \right)\,\left (r\,\frac{\partial}{\partial r}-m\,\omega\,r\,f (r) \right)+\frac{1}{\alpha^2\,r^2}\,\frac{\partial^2}{\partial \phi^2}+\frac{\partial^2}{\partial z^2}\nonumber\\&&+\frac{\partial}{\partial x^2}
-(m+ S(r))^2]\,\Psi (t, r, \phi, z, x)=0,\nonumber\\
&&[-\frac{\partial^2}{\partial t^2}+\frac{\partial^2}{\partial r^2}+\frac{1}{r}\,\frac{\partial}{\partial r}-m\,\omega\,\left (f'(r)+\frac{f (r)}{r} \right)-m^2\,\omega^2\,f^{2} (r)+\frac{1}{\alpha^2\,r^2}\,\frac{\partial^2}{\partial \phi^2}\nonumber\\
&&+\frac{\partial^2}{\partial z^2}+\frac{\partial}{\partial x^2}-(m+ S(r))^2]\,\Psi (t, r, \phi, z, x)=0.
\label{8}
\end{eqnarray}
Since the metric is independent of $t, \phi ,z, x$. One can choose the following ansatz for the function $\Psi$
\begin{equation}
\Psi (t, r, \phi, z, x)=e^{i\,(-E\,t+l\,\phi+k\,z+q\,x)}\,\psi(r),
\label{9}
\end{equation}
where $E$ is the total energy, $l=0,\pm\,1,\pm\,2..$, and $k, q$ are constants.

Substituting the above ansatz into the Eq. (\ref{8}), we get the following equation for $\psi (r)$ :
\begin{eqnarray}
&&[ \frac{d^2}{dr^2} + \frac{1}{r}\,\frac{d}{dr} + E^2-k^2-q^2-\frac{l^2}{\alpha^2\,r^2}-m\,\omega\,\left (f'(r)+\frac{f (r)}{r} \right)\nonumber\\
&&-m^2\,\omega^2\,f^{2}(r)-\left(m+S(r) \right)^2]\,\psi(r)=0.
\label{10}
\end{eqnarray}

We choose the function $f(r)$ a Cornell-type given by \cite{EPJC,ff5,cc9,SZ}
\begin{equation}
f(r)=a\,r+\frac{b}{r}\quad,\quad a, b>0.
\label{11}
\end{equation}

Substituting the function (\ref{11}) and Cornell potential (\ref{6}) into the Eq. (\ref{9}), we obtain the following equation:
\begin{equation}
\left [\frac{d^2}{dr^2} + \frac{1}{r}\,\frac{d}{dr} + \lambda-\Omega^2\,r^2-\frac{j^2}{r^2}-\frac{2\,m\,\eta_{c}}{r}-2\,m\,\eta_{L}\,r \right]\,\psi(r)=0,
\label{12}
\end{equation}
where
\begin{eqnarray}
&&\lambda=E^2-k^2-q^2-m^2-2\,m\,\omega\,a-2\,m^2\,\omega^2\,a\,b-2\,\eta_{L}\,\eta_{c},\nonumber\\
&&\Omega=\sqrt{m^2\,\omega^2\,a^2+\eta^2_{L}},\nonumber\\
&&j=\sqrt{\frac{l^2}{\alpha^2}+m^2\,\omega^2\,b^2+\eta^{2}_{c}}.
\label{13}
\end{eqnarray}
Transforming $\rho=\sqrt{\Omega}\,r$ into the equation (\ref{12}), we get
\begin{equation}
\left [\frac{d^2}{d\rho^2} + \frac{1}{\rho}\,\frac{d}{d\rho} + \zeta-\rho^2-\frac{j^2}{\rho^2}-\frac{\eta}{\rho}-\theta\,\rho \right]\,\psi (\rho)=0,
\label{14}
\end{equation}
where 
\begin{equation}
\zeta=\frac{\lambda}{\Omega}\quad,\quad \eta=\frac{2\,m\,\eta_c}{\sqrt{\Omega}}\quad,\quad \theta=\frac{2\,m\,\eta_L}{\Omega^{\frac{3}{2}}}.
\label{15}
\end{equation}

Let us impose that $\psi (\rho) \rightarrow 0$ when $\rho \rightarrow 0$ and $\rho \rightarrow \infty$. Suppose the possible solution to the Eq. (\ref{14}) is
\begin{equation}
\psi (\rho)=\rho^{j}\,e^{-\frac{1}{2}\,(\rho+\theta)\,\rho}\,H (\rho).
\label{16}
\end{equation}
Substituting the solution Eq. (\ref{16}) into the Eq. (\ref{14}), we obtain
\begin{equation}
H''(\rho)+\left [\frac{\gamma}{\rho}-\theta-2\,\rho \right ]\,H'(\rho)+\left [-\frac{\beta}{\rho}+\Theta \right]\,H (\rho)=0,
\label{17}
\end{equation}
where
\begin{eqnarray}
&&\gamma=1+2\,j,\nonumber\\
&&\Theta=\zeta+\frac{\theta^2}{4}-2\,(1+j),\nonumber\\
&&\beta=\eta+\frac{\theta}{2}\,(1+2\,j).
\label{18}
\end{eqnarray}
Equation (\ref{17}) is the biconfluent Heun's differential equation \cite{ff3,AHEP,bb15,bb16,aa6,EVBL,EVBL2,EPJC, bb7,bb8,bb9,ff2,ff5,ff6,bb41,bb42,bb46,bb47,dd51,dd52} and $H (\rho)$ is the Heun polynomials.

The above equation (\ref{17}) can be solved by the Frobenius method. We consider the power series solution around the origin \cite{bb43}
\begin{equation}
H (\rho)=\sum_{i=0}^{\infty}\,c_{i}\,\rho^{i}
\label{19}
\end{equation}
Substituting the above power series solution into the Eq. (\ref{17}), we obtain the following recurrence relation for the coefficients:
\begin{equation}
c_{n+2}=\frac{1}{(n+2)(n+2+2\,j)}\,\left[\left\{\beta+\theta\,(n+1) \right\}\,c_{n+1}-(\Theta-2\,n)\,c_{n} \right].
\label{20}
\end{equation}
And the various coefficients are
\begin{eqnarray}
&&c_1=\left(\frac{\eta}{\gamma}-\frac{\theta}{2} \right)\,c_0,\nonumber\\
&&c_2=\frac{1}{4\,(1+j)}\,[\left(\beta+\theta \right)\,c_{1}-\Theta\,c_{0}].
\label{21}
\end{eqnarray}

The quantum theory requires that the wave function $\Psi$ must be normalized. The bound state solutions $\psi (\rho)$ can be obtained because there is no divergence of the wave function at $\rho \rightarrow 0$ and $\rho \rightarrow \infty$. Since we have written the function $H (\rho)$ as a power series expansion around the origin in Eq. (\ref{19}). Thereby, bound state solutions can be achieved by imposing that the power series expansion (\ref{19}) becomes a polynomial of degree $n$. Through the recurrence relation (\ref{20}), we can see that the power series expansion (\ref{19}) becomes a polynomial of degree $n$ by imposing two conditions \cite{ff3,AHEP,bb15,bb16,aa6,EVBL,EVBL2,EPJC, bb7,bb8,bb9,ff2,ff5,ff6,bb41,bb42,bb46,bb47}:
\begin{eqnarray}
\Theta&=&2\,n \quad (n=1,2,...),\nonumber\\
c_{n+1}&=&0
\label{23}
\end{eqnarray} 

By analyzing the condition $\Theta=2\,n$, we get expression of the energy eigenvalues $E_{n,l}$:
\begin{eqnarray}
&&\frac{\lambda}{\Omega}+\frac{\theta^2}{4}-2\,(1+j)=2\,n\nonumber\\\Rightarrow
&&E^{2}_{n,l}=k^2+q^2+m^2+2\,\Omega\,\left(n+1+\sqrt{\frac{l^2}{\alpha^2}+m^2\,\omega^2\,b^2+\eta^{2}_{c}} \right)\nonumber\\
&&+2\,m^2\,\omega^2\,a\,b+2\,m\,\omega\,a+2\,\eta_{L}\,\eta_c-\frac{m^2\,\eta^2_{L}}{\Omega^2}.
\label{24}
\end{eqnarray}

We plot graphs of the above energy eigenvalues w. r. t. different parameters. In fig. 1, the energy eigenvalues $E_{1,1}$ against the parameter $\eta_c$. In fig. 2, the energy eigenvalues $E_{1,1}$ against the parameter $\eta_L$. In fig. 3, the energy eigenvalues $E_{1,1}$ against the parameter $M$. In fig. 4, the energy eigenvalues $E_{1,1}$ against the parameter $\omega$. In fig. 5, the energy eigenvalues $E_{1,1}$ against the parameter $\Omega$.

Now we impose additional condition $c_{n+1}=0$ to find the individual energy levels and corresponding wave functions one by one as done in \cite{bb44,bb45}. As example, for $n=1$, we have $\Theta=2$ and $c_2=0$ which implies
\begin{eqnarray}
&&c_1=\frac{2}{\beta+\theta}\,c_0\Rightarrow\left(\frac{\eta}{1+2\,j}-\frac{\theta}{2} \right)=\frac{2}{\beta+\theta}\nonumber\\
&&\Omega^3_{1,l}-\frac{\eta^2}{2\,(1+2\,j)}\Omega^2_{1,l}-\eta\,\theta\,(\frac{1+j}{1+2\,j})\,\Omega_{1,l}-\frac{\theta^2}{8}\,(3+2\,j)=0
\label{25}
\end{eqnarray}
a constraint on the parameter $\Omega_{1,l}$. The relation given in Eq. (\ref{25}) gives the possible values of the parameter $\Omega_{1,l}$ that permit us to construct first degree polynomial to H(x) for $n=1$. Note that its values changes for each quantum number $n$ and $l$, so we have labeled $\Omega \rightarrow \Omega_{n,l}$. Besides, since this parameter is determined by the frequency, hence, the frequency $\omega_{1,l}$ is so adjusted that the Eq. (\ref{25}) can be satisfied, where we have simplified our notation by labeling:
\begin{equation}
\omega_{1,l}=\frac{1}{m\,a}\sqrt{\Omega^2_{1,l}-\eta^2_{L}}.
\label{26}
\end{equation}
It is noteworthy that a third-degree algebraic equation (\ref{25}) has at least one real solution and it is exactly this solution that gives us the allowed values of the frequency for the lowest state of the system, which we do not write because its expression is very long. We can note, from Eq. (\ref{25}) that the possible values of the frequency depend on the quantum numbers and the potential parameter. In addition, for each relativistic energy level, we have a different relation of the magnetic field associated to the Cornell-type potential and quantum numbers of the system $\{l, n \}$. For this reason, we have labeled the parameters $\Omega$ and $\omega$ in Eqs. (\ref{25}) and (\ref{26}).

Therefore, the ground state energy level and corresponding wave-function for $n=1$ are given by
\begin{eqnarray}
&&E^{2}_{1,l}=k^2+q^2+m^2+2\,\Omega_{1,l}\,\left(2+\sqrt{\frac{l^2}{\alpha^2}+m^2\,\omega^2\,b^2+\eta^{2}_{c}} \right)\nonumber\\
&&+2\,m^2\,\omega^2_{1,l}\,a\,b+2\,m\,\omega_{1,l}\,a+2\,\eta_{L}\,\eta_c-\frac{m^2\,\eta^2_{L}}{\Omega^2_{1,l}},\nonumber\\
&&\psi_{1,l}=\rho^{\sqrt{\frac{l^2}{\alpha^2}+m^2\,\omega^2_{1,l}\,b^2+\eta^{2}_{c}}}\,e^{-\frac{1}{2}\,\left (\frac{2\,m\,\eta_L}{\Omega^{\frac{3}{2}}_{1,l}}+\rho \right)\,\rho}\,\left(c_0+c_1\,\rho\right),
\label{27}
\end{eqnarray}
where
\begin{eqnarray}
c_1&=&\frac{1}{\Omega^{\frac{1}{2}}_{1,l}}\,\left [\frac{2\,m\,\eta_c}{\left(1+2\,\sqrt{\frac{l^2}{\alpha^2}+m^2\,\omega^2_{1,l}\,b^2+\eta^{2}_{c}} \right)}-\frac{m\,\eta_L}{\Omega_{1,l}} \right]\,c_0.
\label{28}
\end{eqnarray}
Then, by substituting the real solution of Eq. (\ref{26}) into the Eqs. (\ref{27})-(\ref{28}) it is possible to obtain the allowed values of the relativistic energy for the radial mode $n=1$ of a position dependent mass system. We can see that the lowest energy state defined by the real solution of the algebraic equation given in Eq. (\ref{26}) plus the expression given in Eq. (\ref{27}) is defined by the radial mode $n=1$, instead of $n=0$. This effect arises due to the presence of the Cornell-type potential in the system.

For $\alpha \rightarrow 1$, the relativistic energy eigenvalue (\ref{25}) becomes
\begin{eqnarray}
&&E^{2}_{n,l}=k^2+q^2+m^2+2\,\Omega\,\left(n+1+\sqrt{l^2+m^2\,\omega^2\,b^2+\eta^{2}_{c}} \right)\nonumber\\
&&+2\,m^2\,\omega^2\,a\,b+2\,m\,\omega\,a+2\,\eta_{L}\,\eta_c-\frac{m^2\,\eta^2_{L}}{\Omega^2}.
\label{29}
\end{eqnarray}
Equation (\ref{29}) is the relativistic energy eigenvalue of a scalar particles via the generalized Klein-Gordon oscillator subject to a Cornell-type potential in the Minkowski space-time in the Kaluza-Klein theory.

We discuss bellow a very special case of the above relativistic system.

\vspace{0.3cm}
{\bf Case A}: Considering $\eta_{L}=0$, that is, only Coulomb-type potential $S(r)=\frac{\eta_c}{r}$. 
\vspace{0.3cm}

We want to investigate the effect of Coulomb-type potential on a scalar particle in the background of cosmic string space-time in the Kaluza-Klein theory. In that case, the radial wave-equation Eq. (\ref{12}) becomes
\begin{equation}
\left [\frac{d^2}{dr^2}+\frac{1}{r}\,\frac{d}{dr}+\lambda_0-m^2\,\omega^2\,a^2\,r^2-\frac{j^2}{r^2}-\frac{2\,m\,\eta_{c}}{r} \right]\,\psi(r)=0,
\label{aa1}
\end{equation}
where
\begin{equation}
\lambda_0=E^2-k^2-q^2-m^2-2\,m\,\omega\,a-2\,m^2\,\omega^2\,a\,b
\label{aa2}
\end{equation}

Transforming $\rho=\sqrt{m\,\omega\,a}\,r$ into the Eq. (\ref{aa1}), we get
\begin{equation}
\left [\frac{d^2}{d\rho^2}+\frac{1}{\rho}\,\frac{d}{d\rho}+\frac{\lambda_0}{m\,\omega\,a}-\rho^2-\frac{j^2}{\rho^2}-\frac{2\,m\,\eta_c}{\sqrt{m\,\omega\,a}}\,\frac{1}{\rho} \right]\,\psi(\rho)=0.
\label{aa6}
\end{equation}
Suppose the possible solution to Eq. (\ref{aa6}) is
\begin{equation}
\psi (\rho)=\rho^{j}\,E^{-\frac{\rho^2}{2}}\,H (\rho).
\label{aa7}
\end{equation}
Substituting the solution Eq. (\ref{aa7}) into the Eq. (\ref{aa6}), we obtain
\begin{equation}
H ''(\rho)+\left [\frac{1+2\,j}{\rho}-2\,\rho \right]\, H' (\rho)+\left[-\frac{\tilde{\eta}}{\rho}+\frac{\lambda_0}{m\,\omega\,a}-2\,(1+j) \right]\, H (\rho),
\label{aa8}
\end{equation}
where $\tilde{\eta}=\frac{2\,m\,\eta_c}{\sqrt{m\,\omega\,a}}$. Equation (\ref{aa8}) is the Heun's differential equation \cite{ff3,AHEP,bb15,bb16,aa6,EVBL,EVBL2,EPJC,bb7,bb8,bb9,ff2,ff5,ff6,bb41,bb42,bb46,bb47,dd51,dd52} with $H (\rho)$ is the Heun polynomial.

Substituting the power series solution Eq. (\ref{19}) into the Eq. (\ref{aa8}), we obtain the following recurrence relation for coefficients
\begin{equation}
c_{n+2}=\frac{1}{(n+2)(n+2+2\,j)}\,\left [\tilde{\eta}\,c_{n+1}-\{\frac{\lambda_0}{m\,\omega\,a}-2\,(1+j)-2\,n \}\,c_n \right]
\label{aa9}
\end{equation}
The power series solution becomes a polynomial of degree $n$ provided \cite{ff3,AHEP,bb15,bb16,aa6,EVBL,EVBL2,EPJC, bb7,bb8,bb9,ff2,ff5,ff6,bb41,bb42,bb46,bb47}
\begin{eqnarray}
\frac{\lambda_0}{m\,\omega\,a}-2\,(1+j)&=&2\,n\quad (n=1,2,...)\nonumber\\
c_{n+1}&=&0.
\label{aa10}
\end{eqnarray}

Using the first condition, one will get the following energy eigenvalues of the relativistic system :
\begin{eqnarray}
&&E_{n,l}=\pm\{k^2+q^2+m^2+2\,m\,\omega\,a\,\left(n+2+\sqrt{\frac{l^2}{\alpha^2}+m^2\,\omega^2\,b^2+\eta^{2}_c} \right)\nonumber\\
&&+2\,m^2\,\omega^2\,a\,b\}^{\frac{1}{2}}.
\label{aa3}
\end{eqnarray}

The ground state energy levels and corresponding wave-function for $n=1$ are given by
\begin{eqnarray}
&&E_{1,l}=\pm\{k^2+q^2+m^2+2\,m\,\omega_{1,l}\,a\,\left(3+\sqrt{\frac{l^2}{\alpha^2}+m^2\,\omega^2\,b^2+\eta^{2}_{c}} \right)\nonumber\\
&&+2\,m^2\,\omega^2\,a\,b \}^{\frac{1}{2}},\nonumber\\
&&\psi_{1,l} (\rho)=\rho^{\sqrt{\frac{l^2}{\alpha^2}+m^2\,\omega^2_{1,l}\,b^2+\eta^{2}_{c}}}\,e^{-\frac{\rho^2}{2}}\,\left(c_0+c_1\,\rho \right),
\label{aa4}
\end{eqnarray}
where
\begin{eqnarray}
c_1&=&\frac{2\,m\,\eta_{c}}{\sqrt{m\,\omega_{1,l}\,a}\,\left(1+2\,\sqrt{\frac{l^2}{\alpha^2}+m^2\,\omega^2_{1,l}\,b^2+\eta^{2}_{c}}\right)}\nonumber\\
&=&\left(\frac{2}{1+2\,\sqrt{\frac{l^2}{\alpha^2}+m^2\,\omega^2_{1,l}\,b^2+\eta^{2}_{c}}}\right)^{\frac{1}{2}}\,c_0,\nonumber\\
\omega_{1,l}&=&\frac{2\,m\,\eta^2_{c}}{a\,\left(1+2\,\sqrt{\frac{l^2}{\alpha^2}+m^2\,\omega^2_{1,l}\,b^2+\eta^{2}_c} \right)}.
\label{aa5}
\end{eqnarray}
a constraint on the frequency parameter $\omega_{1,l}$.

\vspace{0.3cm}
{\bf Case B}:
\vspace{0.3cm}

We consider another case corresponds to $a \rightarrow 0$, $b \rightarrow 0$ and $\eta_{L}=0$, that is, a scalar quantum particle in the cosmic string background subject to a Coulomb-type scalar potential within the Kaluza-Klein theory. In that case, from Eq. (\ref{12}) we obtain the following equation:
\begin{equation}
\psi''(r)+\frac{1}{r}\,\psi'(r)+[\tilde{\lambda}-\frac{\tilde{j}^2}{r^2}-\frac{2\,m\,\eta_{c}}{r}]\,\psi(r)=0.
\label{bb1}
\end{equation}
Equation (\ref{bb1}) can be written as
\begin{equation}
\psi''(r)+\frac{1}{r}\,\psi'(r)+\frac{1}{r^2}\,(-\xi_1\,r^2+\xi_2\,r-\xi_3)\,\psi(r)=0,
\label{bb2}
\end{equation}
where
\begin{equation}
\xi_1=-\tilde{\lambda}=-(E^2-k^2-q^2-m^2),\quad \xi_2=-2\,m\,\eta_{c},\quad \xi_3=\tilde{j}^2=\frac{l^2}{\alpha^2}+\eta^2_{c}.
\label{bb3}
\end{equation}

Compairing the Eq (\ref{bb2}) with Equation (\ref{A.1}) in appendix A, we get
\begin{eqnarray}
&&\alpha_1=1,\quad \alpha_2=0,\quad \alpha_3=0,\quad \alpha_4=0,\quad \alpha_5=0,\quad \alpha_6=\xi_1,\nonumber\\
&&\alpha_7=-\xi_2,\quad \alpha_8=\xi_3,\quad \alpha_9=\xi_1,\quad \alpha_{10}=1+2\,\sqrt{\xi_3},\nonumber\\
&&\alpha_{11}=2\,\sqrt{\xi_1},\quad \alpha_{12}=\sqrt{\xi_3},\quad \alpha_{13}=-\sqrt{\xi_1}.
\label{bb6}
\end{eqnarray}

The energy eigenvalues using Eqs. (\ref{bb3})-(\ref{bb6}) into the Eq. (\ref{A.8}) in appendix A is given by
\begin{equation}
E_{n,l}=\pm\,m\,\sqrt{1-\frac{\eta^2_{c}}{(n+\sqrt{\frac{l^2}{\alpha^2}+\eta^{2}_{c}}+\frac{1}{2})^2}+\frac{k^2}{m^2}+\frac{q^2}{m^2}},
\label{bb4}
\end{equation}
where $n=0,1,2,..$ is the quantum number associated with the radial modes, $l=0,\pm\,1,\pm\,2,.$ are the quantum number associated with the angular momentum operator, $k$ and $q$ are arbitrary constants. Equation (\ref{bb4}) corresponds to the relativistic energy eigenvalues of a free-scalar particle subject to a Coulomb-type scalar potential in the background of cosmic string within the Kaluza-Klein theory.

The corresponding radial wave-function is given by
\begin{eqnarray}
\psi_{n,l} (r)&=&|N|\,r^{\frac{\tilde{j}}{2}}\,{\sf e}^{-\frac{r}{2}}\,{\sf L}^{(\tilde{j})}_{n} (r)\nonumber\\
&=&|N|\,r^{\frac{1}{2}\sqrt{\frac{l^2}{\alpha^2}+\eta^2_{c}}}\,{\sf e}^{-\frac{r}{2}}\,{\sf L}^{(\sqrt{\frac{l^2}{\alpha^2}+\eta^2_{c}})}_{n} (r).
\label{bb7}
\end{eqnarray}
Here $|N|$ is the normalization constant and ${\sf L}^{(\sqrt{\frac{l^2}{\alpha^2}+\eta^2_{c}})}_{n} (r) $ is the generalized Laguerre polynomial.

For $\alpha \rightarrow 1$, the relativistic energy eigenvalues Eq. (\ref{bb4}) becomes
\begin{equation}
E_{n,l}=\pm\,m\,\sqrt{1-\frac{\eta^2_{c}}{(n+\sqrt{l^2+\eta^{2}_{c}}+\frac{1}{2})^2}+\frac{k^2}{m^2}+\frac{q^2}{m^2}}.
\label{bb5}
\end{equation}
Equation (\ref{bb5}) correspond to the relativistic energy eigenvalue of a scalar particle subject to a Coulomb-type scalar potential in the Minkowski space-time within the Kaluza-Klein theory.

\section{Generalized Klein-Gordon oscillator in the magnetic cosmic string with a Coulomb-type potential in Kaluza-Klein theory}

Let us consider the quantum dynamics of a particle moving in the magnetic cosmic string background. In the Kaluza-Klein theory \cite{bb12,bb13,bb28}, the corresponding metrics with Aharonov-Bohm magnetic flux $\Phi$ passing along the symmetry axis of the string assumes the following form
\begin{equation}
ds^2=-dt^2+dr^2+\alpha^2\,r^2\,d\phi^2+dz^2+(dx+\frac{\Phi}{2\,\pi}\,d\phi)^2
\label{30}
\end{equation}
with cylindrical coordinates are used. The quantum dynamics is described by the equation (\ref{4}) with the following change in the inverse matrix tensor $g^{\mu\nu}$,
\begin{equation}
g^{\mu\nu}=\left (\begin{array}{lllll}
-1 
& 0 & \quad 0 & 0 & \quad 0 \\
\quad 0 & 1 & \quad 0 & 0 & \quad 0 \\
\quad 0 & 0 & \quad \frac{1}{\alpha^2\,r^2} & 0 & -\frac{\Phi}{2\,\pi\,\alpha^2\,r^2} \\
\quad 0 & 0 & \quad 0 & 1 & \quad 0 \\
\quad 0 & 0 & -\frac{\Phi}{2\,\pi\,\alpha^2\,r^2} & 0 & 1+\frac{\Phi^2}{4\,\pi^2\,\alpha^2\,r^2}
\end{array} \right).
\label{31}
\end{equation}
By considering the line element (\ref{30}) into the Eq. (\ref{4}), we obtain the following differential equation :
\begin{eqnarray}
&&[-\partial_{t}^2+\partial_{r}^2+\frac{1}{r}\,\partial_{r}+\frac{1}{\alpha^2\,r^2}\,(\partial_{\phi}-\frac{\Phi}{2\,\pi}\,\partial_{x})^2+\partial_{z}^2+\partial_{x}^2\nonumber\\
&&-m\,\omega\,\left(f' (r)+\frac{f(r)}{r} \right)-m^2\,\omega^2\,f^{2}(r)-\left(m + S(r) \right)^2]\,\Psi=0.
\label{32}
\end{eqnarray}
Since the space-time is independent of $t, \phi, z, x$, substituting the ansatz (\ref{9}) into the Eq. (\ref{32}), we get the following equation :
\begin{eqnarray}
&&\psi ''(r)+\frac{1}{r}\,\psi'(r)+[E^2-k^2-q^2-\frac{l^2_{eff}}{r^2}-m\,\omega\,\left(f'(r)+\frac{f(r)}{r} \right)\nonumber\\
&-&m^2\,\omega^2\,f^{2}(r)-\left(m+S (r) \right)^2]\,\psi (r)=0,
\label{33}
\end{eqnarray}
where the effective angular quantum number
\begin{equation}
l_{eff}=\frac{1}{\alpha}\,(l-\frac{q\,\Phi}{2\,\pi}).
\label{34}
\end{equation}

Substituting the function (\ref{11}) into the Eq. (\ref{33}) and using Coulomb-type potential (\ref{5}), the radial wave-equation becomes
\begin{equation}
\left [\frac{d^2}{dr^2} + \frac{1}{r}\,\frac{d}{dr} + \lambda_0-m^2\,\omega^2\,a^2\,r^2-\frac{\chi^2}{r^2}-\frac{2\,m\,\eta_{c}}{r} \right]\,\psi(r)=0,
\label{35}
\end{equation}
where
\begin{eqnarray}
&&\lambda_0=E^2-k^2-q^2-m^2-2\,m\,\omega\,a-2\,m^2\,\omega^2\,a\,b,\nonumber\\
&&\chi=\sqrt{\frac{(l-\frac{q\,\Phi}{2\,\pi})^2}{\alpha^2}+m^2\,\omega^2\,b^2+\eta^{2}_{c}}.
\label{36}
\end{eqnarray}
Transforming $\rho=\sqrt{m\,\omega\,a}\,r$ into the equation (\ref{35}), we get
\begin{equation}
\left [\frac{d^2}{d\rho^2}+\frac{1}{\rho}\,\frac{d}{d\rho}+ \frac{\lambda_0}{m\,\omega\,a}-\rho^2-\frac{\chi^2}{\rho^2}-\frac{\tilde{\eta}}{\rho} \right]\,\psi (\rho)=0,
\label{37}
\end{equation}
where $\tilde{\eta}=\frac{2\,m\,\eta_c}{\sqrt{m\,\omega\,a}}$.

Suppose the possible solution to Eq. (\ref{37}) is
\begin{equation}
\psi (\rho)=\rho^{\chi}\,e^{-\frac{\rho^2}{2}}\,H (\rho)
\label{42}
\end{equation}
Substituting the solution Eq. (\ref{42}) into the Eq. (\ref{37}), we obtain
\begin{equation}
H'' (\rho)+\left[\frac{1+2\,\chi}{\rho}-2\,\rho \right]\,H' (\rho)+\left [-\frac{\tilde{\eta}}{\rho}+\frac{\lambda_0}{m\,\omega\,a}-2\,(1+\chi) \right]\,H (\rho).
\label{43}
\end{equation}
Equation (\ref{43}) is the second order Heun's differential equation \cite{ff3,AHEP,bb15,bb16,aa6,EVBL,EVBL2,EPJC, bb7,bb8,bb9,ff2,ff5,ff6,bb41,bb42,bb46,bb47,dd51,dd52} with $H (\rho)$ is the Heun polynomial.

Substituting the power series solution Eq. (\ref{19}) into the Eq. (\ref{43}), we obtain the following recurrence relation for the coefficients:
\begin{equation}
c_{n+2}=\frac{1}{(n+2)\,(n+2+2\,\chi)}\,\left[\tilde{\eta}\,c_{n+1}-\left\{ \frac{\lambda_0}{m\,\omega\,a}-2-2\,\chi-2\,n \right\}\,c_n \right].
\label{44}
\end{equation}
The power series becomes a polynomial of degree $n$ by imposing the following conditions \cite{ff3,AHEP,bb15,bb16, aa6,EVBL,EVBL2,EPJC,bb7,bb8,bb9,ff2,ff5,ff6,bb41,bb42,bb46,bb47}
\begin{equation}
c_{n+1}=0\quad,\quad \frac{\lambda_0}{m\,\omega\,a}-2-2\,\chi=2\,n\quad (n=1,2,...)
\label{45}
\end{equation}

By analyzing the second condition, we get the following energy eigenvalues $E_{n,l}$:
\begin{eqnarray}
&&E^{2}_{n,l}=k^2+q^2+m^2+2\,m\,\omega\,a\,\left(n+2+\sqrt{\frac{(l-\frac{q\,\Phi}{2\,\pi})^2}{\alpha^2}+m^2\,\omega^2\,b^2+\eta^{2}_c}\right)\nonumber\\
&&+2\,m^2\,\omega^2\,a\,b.
\label{38}
\end{eqnarray}
Equation (\ref{38}) is the energy eigenvalues of a generalized Klein-Gordon oscillator in the magnetic cosmic string with a Coulomb-type scalar potential in the Kaluza-Klein theory. Observed that the relativistic energy eigenvalues Eq. (\ref{38}) depend on the Aharonov-Bohm geometric quantum phase \cite{bb50}. Thus, we have that $E_{n, l} (\Phi+\Phi_0)=E_{n, l \mp \tau} (\Phi)$ where, $\Phi_0=\pm\,\frac{2\,\pi}{q}\,\tau$ with $\tau=0,1,2..$. This dependence of the relativistic energy eigenvalue on the geometric quantum phase $\Phi$ gives rise to a relativistic analogue of the Aharonov-Bohm effect for bound states \cite{ff3,bb15,bb28,bb39,bb40,bb50}.

We plot graphs of the above energy eigenvalues w. r. t. different parameters. In fig. 6, the energy eigenvalues $E_{1,1}$ against the parameter $\eta_c$. In fig. 7, the energy eigenvalues $E_{1,1}$ against the parameter $M$. In fig. 8, the energy eigenvalues $E_{1,1}$ against the parameter $\omega$. In fig. 9, the energy eigenvalues $E_{1,1}$ against the parameter $\Phi$. 

The ground state energy levels and corresponding wave-function for $n=1$ are given by
\begin{eqnarray}
&&E^{2}_{1,l}=k^2+q^2+m^2+2\,m\,\omega_{1,l}\,a\,\left(3+\sqrt{\frac{(l-\frac{q\,\Phi}{2\,\pi})^2}{\alpha^2}+m^2\,\omega^2\,b^2+\eta^{2}_c} \right)\nonumber\\
&&+2\,m^2\,\omega^2_{1,l}\,a\,b\quad,\nonumber\\
&&\psi_{1,l} (\rho)=\rho^{\sqrt{\frac{(l-\frac{q\,\Phi}{2\,\pi})^2}{\alpha^2}+m^2\,\omega^2_{1,l}\,b^2+\eta^{2}_c}}\,e^{-\frac{\rho^2}{2}}\,\left(c_0+c_1\,\rho \right),
\label{39}
\end{eqnarray}
where
\begin{eqnarray}
c_1&=&\frac{2\,m\,\eta_{c}}{\sqrt{m\,\omega_{1,l}\,a}\,(1+2\,\sqrt{\frac{(l-\frac{q\,\Phi}{2\,\pi})^2}{\alpha^2}+m^2\,\omega^2_{1,l}\,b^2+\eta^{2}_{c}})}\nonumber\\
&=&\left(\frac{2}{1+2\,\sqrt{\frac{(l-\frac{q\,\Phi}{2\,\pi})^2}{\alpha^2}+m^2\,\omega^2_{1,l}\,b^2+\eta^{2}_c}}\right)^{\frac{1}{2}}\,c_0. \nonumber\\
\omega_{1,l}&=&\frac{2\,m\,\eta^2_{c}}{a\,\left (1+2\,\sqrt{\frac{(l-\frac{q\,\Phi}{2\,\pi})^2}{\alpha^2}+m^2\,\omega^2_{1,l}\,b^2+\eta^{2}_c}\right)}
\label{40}
\end{eqnarray}
a constraint on the physical parameter $\omega_{1,l}$. 

Equation Eq. (\ref{39}) is the ground states energy eigenvalues and corresponding eigenfunctions of a generalized Klein-Gordon oscillator in the presence of Coulomb-type scalar potential in a magnetic cosmic string space-time in the Kaluza-Klein theory.

For $\alpha \rightarrow 1$, the energy eigenvalues (\ref{38}) becomes
\begin{eqnarray}
&&E^{2}_{n,l}=k^2+m^2+q^2+2\,m\,\omega\,a\,\left(n+2+\sqrt{(l-\frac{q\,\Phi}{2\,\pi})^2+m^2\,\omega^2\,b^2+\eta^{2}_{c}}\right)\nonumber\\
&&+2\,m^2\,\omega^2\,a\,b.
\label{41}
\end{eqnarray}
Equation (\ref{41}) is the relativistic energy eigenvalue of the generalized Klein-Gordon oscillator field with a Coulomb-type scalar potential with a magnetic flux in the Kaluza-Klein theory. Observed that the relativistic energy eigenvalue Eq. (\ref{41}) depend on the geometric quantum phase \cite{bb50}. Thus, we have that $E_{n,l} (\Phi+\Phi_0)=E_{n,l \mp \tau} (\Phi)$ where, $\Phi_0=\pm\,\frac{2\,\pi}{q}\,\tau$ with $\tau=0,1,2..$. This dependence of the relativistic energy eigenvalue on the geometric quantum phase gives rise to an analogous effect to Aharonov-Bohm effect for bound states \cite{ff3,bb15,bb28,bb39,bb40,bb50}.

\vspace{0.3cm}
{\bf Case A}:
\vspace{0.3cm}

We discuss below a special case corresponds to $b \rightarrow 0$, $a \rightarrow 0$, that is, a scalar quantum particle in a magnetic cosmic string background subject to a Coulomb-type scalar potential in the Kaluza-Klein theory. In that case, from Eq. (\ref{35}) we obtain the following equation:
\begin{equation}
\psi''(r)+\frac{1}{r}\,\psi'(r)+[\tilde{\lambda}-\frac{\tilde{\chi}^2}{r^2}-\frac{2\,m\,\eta_{c}}{r}]\,\psi(r)=0,
\label{cc1}
\end{equation}
where
\begin{eqnarray}
&&\tilde{\lambda}=E^2-k^2-q^2-m^2,\nonumber\\
&&\tilde{\chi}_0=\sqrt{\frac{(l-\frac{q\,\Phi}{2\,\pi})^2}{\alpha^2}+\eta^{2}_c}.
\label{cc2}
\end{eqnarray}
The above Eq. (\ref{cc1}) can be written as 
\begin{equation}
\psi''(r)+\frac{1}{r}\,\psi'(r)+\frac{1}{r^2}\,\left(-\xi_1\,r^2+\xi_2\,r-\xi_3 \right)\,\psi(r)=0,
\label{cc3}
\end{equation}
where
\begin{equation}
\xi_1=-\tilde{\lambda}\quad,\quad \xi_2=-2\,m\,\eta_{c}\quad,\quad \xi_3=\tilde{\chi}^{2}_0.
\label{Bakke}
\end{equation}

Following the similar technique as done earlier, we get the following energy eigenvalues $E_{n,l}$: 
\begin{equation}
E_{n,l}=\pm\,m\,\sqrt{1-\frac{\eta^2_{c}}{\left (n+\sqrt{\frac{1}{\alpha^2}\,(l-\frac{q\,\Phi}{2\,\pi})^2+\eta^{2}_{c}}+\frac{1}{2}\right)^2}+\frac{k^2}{m^2}+\frac{q^2}{m^2}},
\label{cc5}
\end{equation}
where $n=0,1,2,..$ is the quantum number associated with radial modes, $l=0,\pm\,1,\pm\,2,....$ are the quantum number associated with the angular momentum, $k$ and $q$ are constants. Equation (\ref{cc5}) corresponds to the relativistic energy levels for a free-scalar particle subject to Coulomb-type scalar potential in the background of magnetic cosmic string in a Kaluza-Klein theory.

The radial wave-function is given by
\begin{eqnarray}
&&\psi_{n,l} (r)=|N|\,r^{\frac{\tilde{\chi}_0}{2}}\,{\sf e}^{-\frac{r}{2}}\,{\sf L}^{(\tilde{j})}_{n} (r)\nonumber\\
&=&|N|\,r^{\frac{1}{2},\sqrt{\frac{(l-\frac{q\,\Phi}{2\,\pi})^2}{\alpha^2}+\eta^{2}_c}}\,{\sf e}^{-\frac{r}{2}}\,{\sf L}^{(\sqrt{\frac{(l-\frac{q\,\Phi}{2\,\pi})^2}{\alpha^2}+\eta^{2}_c})}_{n} (r).
\label{cc7}
\end{eqnarray}
Here $|N|$ is the normalization constant and ${\sf L}^{(\sqrt{\frac{(l-\frac{q\,\Phi}{2\,\pi})^2}{\alpha^2}+\eta^{2}_c})}_{n} (r) $ is the generalized Laguerre polynomial.

For $\alpha \rightarrow 1$, the energy eigenvalues (\ref{cc5}) becomes
\begin{equation}
E_{n,l}=\pm\,m\,\sqrt{1-\frac{\eta^2_{c}}{\left (n+\sqrt{(l-\frac{q\,\Phi}{2\,\pi})^2+k^{2}_{c}}+\frac{1}{2}\right)^2}+\frac{k^2}{m^2}+\frac{q^2}{m^2}},
\label{cc6}
\end{equation}
which is similar to the energy eigenvalue obtained in \cite{bb16} (see Eq. (12) in \cite{bb16}). Thus we can see that the cosmic string $\alpha$ modify the relativistic energy eigenvalue (\ref{cc5}) in comparison to those results obtained in \cite{bb16}.

Observe that the relativistic energy eigenvalues Eq. (\ref{cc5}) depend on the cosmic string parameter $\alpha$, the magnetic quantum flux $\Phi$, and potential parameter $\eta_c$. We can see that $E_{n, l} (\Phi+\Phi_0)=E_{n, l \mp \tau} (\Phi)$ where, $\Phi_0=\pm\,\frac{2\,\pi}{q}\,\tau$ with $\tau=0,1,..$. This dependence of the relativistic energy eigenvalues on the geometric quantum phase gives rise to a relativistic analogue of the Aharonov-Bohm effect for bound states \cite{ff3,bb15,bb28,bb39,bb40,bb50}.

\subsection{Persistent currents of the Relativistic System}

By following \cite{NB,WCT,LD}, the expression for the total persistent currents is given by 
\begin{equation}
I=\sum_{n,l}\,I_{n,l},
\label{dd1}
\end{equation}
where
\begin{equation}
I_{n,l}=-\frac{\partial E_{n,l}}{\partial \Phi}
\label{dd2}
\end{equation}
is called the Byers-Yang relation \cite{NB}. 

Therefore, the persistent current that arises in this relativistic system using Eq. (\ref{38}) is given by
\begin{eqnarray}
&&I_{n,l}=-\frac{\partial E_{n,l}}{\partial \Phi}\nonumber\\
&&=\mp\frac{m\,\omega\, a\,(\frac{\partial\,\chi}{\partial\,\Phi})}{\sqrt{k^2+q^2+m^2+2m^2\omega^2 a b+2 m \omega a\left(n+2+\sqrt{\frac{(l-\frac{q\,\Phi}{2\,\pi})^2}{\alpha^2}+m^2\omega^2 b^2+\eta^{2}_c}\right)}},\quad\quad
\label{bb55}
\end{eqnarray}
where
\begin{eqnarray}
\frac{\partial\,\chi}{\partial\,\Phi}=-\frac{q\,(l-\frac{q\,\Phi}{2\,\pi})}{2\,\alpha^2\,\pi\,\sqrt{\frac{(l-\frac{q\,\Phi}{2\,\pi})^2}{\alpha^2}+m^2\,\omega^2\,b^2+\eta^{2}_c}}.
\label{dd4}
\end{eqnarray}

Similarly, for the relativistic system discussed in {\bf case A} in this section, this current using Eq. (\ref{cc5}) is given by
\begin{eqnarray}
I_{n,l}&=&\pm\,\frac{m\,q\,\eta^{2}_c\,(l-\frac{q\,\Phi}{2\,\pi})}{2\,\pi\,\alpha^2\,\left(n+\frac{1}{2}+\sqrt{\frac{(l-\frac{q\,\Phi}{2\,\pi})^2}{\alpha^2}+\eta^2_{c}} \right)^3\,\sqrt{\frac{(l-\frac{q\,\Phi}{2\,\pi})^2}{\alpha^2}+\eta^2_{c}}}\nonumber\\
&&\times\frac{1}{\sqrt{1-\frac{\eta^2_{c}}{\left(n+\sqrt{\frac{1}{\alpha^2}\,(l-\frac{q\,\Phi}{2\,\pi})^2+\eta^{2}_c}+\frac{1}{2}\right)^2}+\frac{k^2}{m^2}+\frac{q^2}{m^2}}}.
\label{dd5}
\end{eqnarray}
For $\alpha \rightarrow 1$, the persistent currents expression given by Eq. (\ref{dd5}) reduces to the result obtained in Ref. \cite{bb16}. Thus we can see that the presence of the cosmic string parameter modify the persistent currents Eq. (\ref{dd5}) in comparison to those results in Ref. \cite{bb16}.

By introducing a magnetic flux through the line element of the cosmic string space-time in five dimensions, we see that the relativistic energy eigenvalue Eq. (\ref{38}) depend on the geometric quantum phase \cite{bb50} which gives rise to a relativistic analogue of the Aharonov-Bohm effect for bound states \cite{ff3,bb15,bb28,bb39,bb40,bb50}. Moreover, this dependence of the relativistic energy eigenvalues on the geometric quantum phase has yielded persistent currents in this relativistic quantum system.

\section{Conclusions}

In Ref. \cite{bb16}, Aharonov-Bohm effects for bound states of a relativistic scalar particle by solving the Klein-Gordon equation subject to a Coulomb-type potential in the Minkowski space-time within the Kaluza-Klein theory were studied. They obtained the relativistic bound states solutions and calculated the persistent currents. In Ref. \cite{bb14}, it is shown that the cosmic string space-time and the magnetic cosmic string space-time can have analogue in five dimensions. In Ref. \cite{bb28}, quantum mechanics of a scalar particle in the background of a chiral cosmic string using the Kaluza-Klein theory was studied. They shown that the wave functions, the phase shifts, and scattering amplitudes associated with the particle depend on the global features of those space-times. These dependence represent the gravitational analogues of the well-known Aharonov-Bohm effect. In addition, they discussed the Landau levels in the presence of a cosmic string within the framework of Kaluza-Klein theory. In Ref. \cite{aa6}, the Klein-Gordon oscillator on the curved background within the Kaluza-Klein theory were studied. The problem of the interaction between particles coupled harmonically with topological defects in the Kaluza-Klein theory were studied. They considered a series of topological defects and then treated the Klein-Gordon oscillator coupled to this background, and obtained the energy eigenvalue and corresponding eigenfunctions in this cases. They have shown that the energy eigenvalue depend on the global parameters characterizing these space-times. In Ref. \cite{EVBL}, a scalar particle with position-dependent mass subject to a uniform magnetic field and a quantum magnetic flux, both coming from the background which is governed by the Kaluza-Klein theory were investigated. They inserted a Cornell-type scalar potential into this relativistic systems and determined the relativistic energy eigenvalue of the system in this background of extra dimension. They analyzed particular cases of this system and a quantum effect were observed: the dependence of the magnetic field on the quantum numbers of the solutions. In Ref. \cite{EPJC}, the relativistic quantum dynamics of a scalar particle subject to linear potential on the curved background within the Kaluza-Klein theory was studied. We have solved the generalized Klein-Gordon oscillator in the cosmic string and magnetic cosmic string space-time with a linear potential within the Kaluza-Klein theory. We have shown that the energy eigenvalues obtained there depend on the global parameters characterizing these space-times and the gravitational analogue to the Aharonov-Bohm effect for bound states \cite{ff3,bb15,bb28,bb39,bb40,bb50} of a scalar particle was analyzed.
 
In this work, we have investigated the relativistic quantum dynamics of a scalar particle interacting with gravitational fields produced by topological defects via the Klein-Gordon oscillator of the Klein-Gordon equation in the presence of cosmic string and magnetic cosmic string within the Kaluza-Klein theory with scalar potential. We have determined the manner in which the non-trivial topology due to the topological defects and a quantum magnetic flux modifies the energy spectrum and wave-functions of a scalar particle. We then have studied the quantum dynamics of a scalar particle interacting with fields by introducing a magnetic flux through the line element of a cosmic string space-time using the five-dimensional version of the General Relativity. The quantum dynamics in the usual as well as magnetic cosmic string cases allow us to obtain the energy eigenvalues and corresponding wave-functions that depend on the external parameters characterize the background space-time, a result known by gravitational analogue of the well studied Aharonov-Bohm effect. 

In {\it section 2}, we have chosen a Cornell-type function $f(r)=a\,r+\frac{b}{r}$ and  Cornell-type potential $S(r)=\eta_{L}\,r+\frac{\eta_c}{r}$ into the relativistic systems. We have solved the generalized Klein-Gordon oscillator in the cosmic string background within the Kaluza-Klein theory and obtained the energy eigenvalues Eq. (\ref{24}). We have plotted graphs of the energy eigenvalues Eq. (\ref{24}) w. r. t. different parameters by figs. 1--5. By imposing the additional recurrence condition $c_{n+1}=0$ on the relativistic eigenvalue problem, for example $n=1$, we have obtained the ground state energy levels and wave-functions by Eqs. (\ref{27})--(\ref{28}). We have discussed a special case corresponds to $\eta_{L} \rightarrow 0$ and obtained the relativistic energy eigenvalues Eq. (\ref{aa3}) of a generalized Klein-Gordon oscillator in the cosmic string space-time within the Kaluza-Klein theory. We have also obtained the relativistic energy eigenvalues Eq. (\ref{bb4}) of a free-scalar particle by solving the Klein-Gordon equation with a Coulomb-type scalar potential in the background of cosmic string space-time in the Kaluza-Klein theory. 

In {\it section 3}, we have studied the relativistic quantum dynamics of a scalar particle in the background of magnetic cosmic string in the Kaluza-Klein theory with a scalar potential. By choosing the same function $f(r)=a\,r+\frac{b}{r}$ and a Coulomb-type scalar potential $S(r)=\frac{\eta_c}{r}$, we have solved the radial wave-equation in the considered system and obtained the bound states energy eigenvalues Eq. (\ref{38}). We have plotted graphs of the energy eigenvalues Eq. (\ref{38}) w. r. t. different parameters by figs. 6--9. Subsequently, the ground state energy levels Eq. (\ref{39}) and corresponding wave-functions Eq. (\ref{40}) for the radial mode $n=1$ by imposing the additional condition $c_{n+1}=0$ on the eigenvalue problem is obtained. Furthermore, a special case corresponds to $a\rightarrow 0$, $b\rightarrow 0$ is discussed and obtained the relativistic energy eigenvalues Eq. (\ref{cc5}) of a scalar particle by solving the Klein-Gordon equation with a Coulomb-type scalar potential in the magnetic cosmic string space-time in the Kaluza-Klein theory. For $\alpha \rightarrow 1$, we have seen that the energy eigenvalues Eq. (\ref{cc5}) reduces to the result obtained in Ref. \cite{bb16}. As there is an effective angular momentum quantum number, $l \rightarrow l_{eff}=\frac{1}{\alpha}\,(l-\frac{q\,\Phi}{2\pi})$, thus the relativistic energy eigenvalues Eqs. (\ref{38}) and (\ref{cc5}) depend on the geometric quantum phase \cite{bb50}. Hence, we have that $E_{n, l} (\Phi+\Phi_0)=E_{n, l \mp \tau} (\Phi)$ where, $\Phi_0=\pm\,\frac{2\,\pi}{q}\,\tau$ with $\tau=0,1,2,.$. This dependence of the relativistic energy eigenvalues on the geometric quantum phase gives rise to a relativistic analogue of the Aharonov-Bohm effect for bound states \cite{bb15,bb39,bb40,bb50}. Finally, we have obtained the persistent currents by Eqs. (\ref{bb55})--(\ref{dd5}) for this relativistic quantum system because of the dependence of the relativistic energy eigenvalues on the geometric quantum phase.

So in this paper, we have shown some results which are in addition to those results obtained in Refs. \cite{bb28,bb15,bb16,aa6,EVBL,EVBL2,EPJC} presents many interesting effects.

\section*{Data Availability}

No data has been used to prepare this paper.

\section*{Conflict of Interest}

Author declares that there is no conflict of interest regarding publication this paper.

\section*{Acknowledgement}

Author sincerely acknowledge the anonymous kind referee(s) for their valuable comments and suggestions and thanks the editor.

\section*{Appendix A : Brief review of the Nikiforov-Uvarov (NU) method}

\setcounter{equation}{0}
\renewcommand{\theequation}{A.\arabic{equation}}

The Nikiforov-Uvarov method is helpful in order to find eigenvalues and eigenfunctions of the Schr\"{o}dinger like equation, as well as other second-order differential equations of physical interest. According to this method, the eigenfunctions of a second-order differential equation \cite{bb49}
\begin{equation}
\frac{d^2 \psi (s)}{ds^2}+\frac{(\alpha_1-\alpha_2\,s)}{s\,(1-\alpha_3\,s)}\,\frac{d \psi (s)}{ds}+\frac{(-\xi_1\,s^2+\xi_2\,s-\xi_3)}{s^2\,(1-\alpha_3\,s)^2}\,\psi (s)=0.
\label{A.1}
\end{equation}
are given by 
\begin{equation}
\psi (s)=s^{\alpha_{12}}\,(1-\alpha_3\,s)^{-\alpha_{12}-\frac{\alpha_{13}}{\alpha_3}}\,P^{(\alpha_{10}-1,\frac{\alpha_{11}}{\alpha_3}-\alpha_{10}-1)}_{n}\,(1-2\,\alpha_3\,s).
\label{A.2}
\end{equation}
And that the energy eigenvalues equation
\begin{eqnarray}
&&\alpha_2\,n-(2\,n+1)\,\alpha_5+(2\,n+1)\,(\sqrt{\alpha_9}+\alpha_3\,\sqrt{\alpha_8})+n\,(n-1)\,\alpha_3+\alpha_7\nonumber\\
&&+2\,\alpha_3\,\alpha_8+2\,\sqrt{\alpha_8\,\alpha_9}=0.
\label{A.3}
\end{eqnarray}
The parameters $\alpha_4,\ldots,\alpha_{13}$ are obtained from the six parameters $\alpha_1,\ldots,\alpha_3$ and $\xi_1,\ldots,\xi_3$ as follows:
\begin{eqnarray}
&&\alpha_4=\frac{1}{2}\,(1-\alpha_1)\quad,\quad \alpha_5=\frac{1}{2}\,(\alpha_2-2\,\alpha_3),\nonumber\\
&&\alpha_6=\alpha^2_{5}+\xi_1\quad,\quad \alpha_7=2\,\alpha_4\,\alpha_{5}-\xi_2,\nonumber\\
&&\alpha_8=\alpha^2_{4}+\xi_3\quad,\quad \alpha_9=\alpha_6+\alpha_3\,\alpha_7+\alpha^{2}_3\,\alpha_8,\nonumber\\
&&\alpha_{10}=\alpha_1+2\,\alpha_4+2\,\sqrt{\alpha_8}\quad,\quad \alpha_{11}=\alpha_2-2\,\alpha_5+2\,(\sqrt{\alpha_9}+\alpha_3\,\sqrt{\alpha_8}),\nonumber\\
&&\alpha_{12}=\alpha_4+\sqrt{\alpha_8}\quad,\quad \alpha_{13}=\alpha_5-(\sqrt{\alpha_9}+\alpha_3\,\sqrt{\alpha_8}).
\label{A.4}
\end{eqnarray}

A special case where $\alpha_3=0$, as in our case, we find
\begin{equation}
\lim_{\alpha_3\rightarrow 0} P^{(\alpha_{10}-1,\frac{\alpha_{11}}{\alpha_3}-\alpha_{10}-1)}_{n}\,(1-2\,\alpha_3\,s)=L^{\alpha_{10}-1}_{n} (\alpha_{11}\,s),
\label{A.5}
\end{equation}
and 
\begin{equation}
\lim_{\alpha_3\rightarrow 0} (1-\alpha_3\,s)^{-\alpha_{12}-\frac{\alpha_{13}}{\alpha_3}}=e^{\alpha_{13}\,s}.
\label{A.6}
\end{equation}
Therefore the wave-function from (\ref{A.2}) becomes
\begin{equation}
\psi (s)=s^{\alpha_{12}}\,e^{\alpha_{13}\,s}\,L^{\alpha_{10}-1}_{n} (\alpha_{11}\,s),
\label{A.7}
\end{equation}
where $L^{(\alpha)}_{n} (x)$ denotes the generalized Laguerre polynomial. 

The energy eigenvalues equation reduces to 
\begin{equation}
n\,\alpha_2-(2\,n+1)\,\alpha_5+(2\,n+1)\,\sqrt{\alpha_9}+\alpha_7+2\,\sqrt{\alpha_8\,\alpha_9}=0.
\label{A.8}
\end{equation}
Noted that the simple Laguerre polynomial is the special case $\alpha=0$ of the generalized Laguerre polynomial:
\begin{equation}
L^{(0)}_{n} (x)=L_{n} (x).
\label{A.9}
\end{equation}

\pagebreak 



\begin{figure}
\centering
\includegraphics[width=3.0in,height=2.0in]{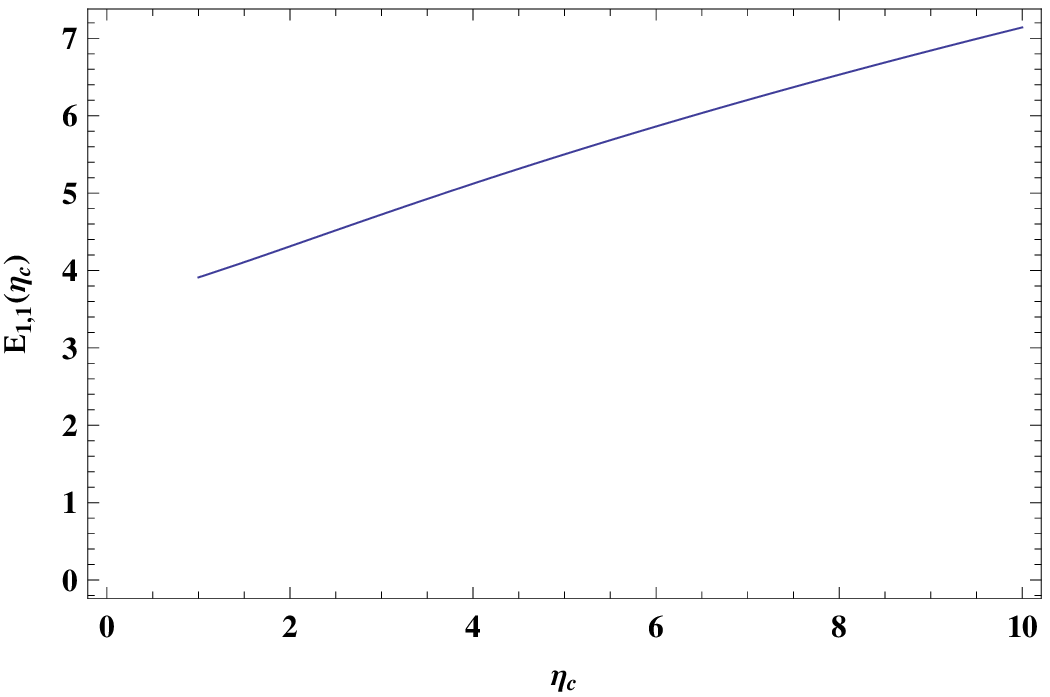}
\caption{$n=l=k=M=q=a=b=\eta_L=1$, $\alpha=0.5$, $\omega=0.5$}
\end{figure}
\begin{figure}
\centering
\includegraphics[width=3.0in,height=2.0in]{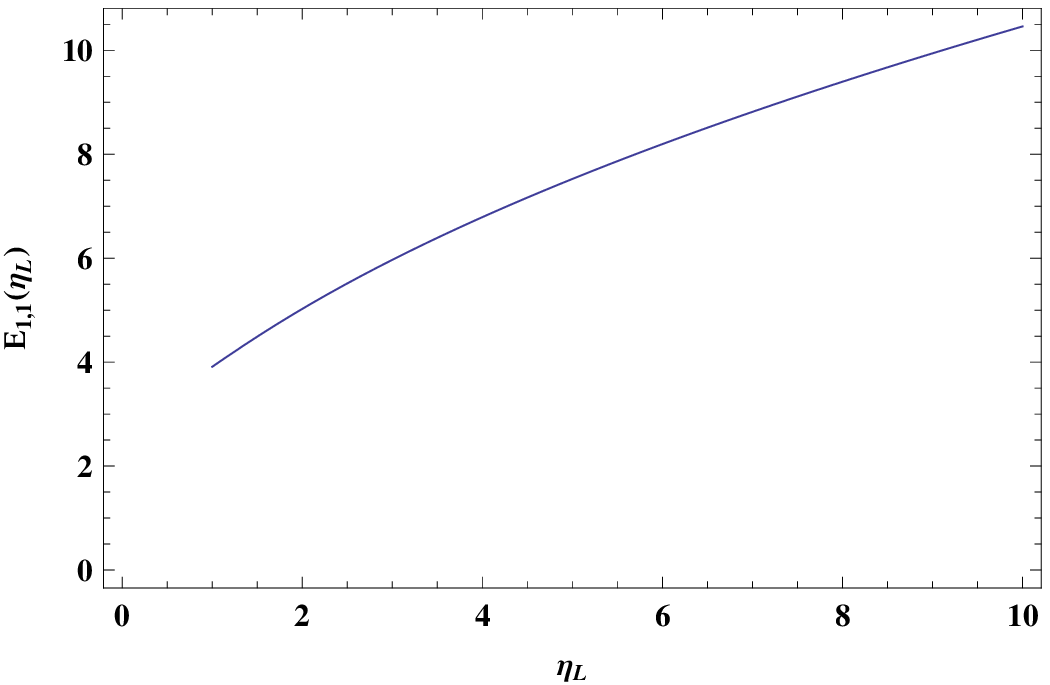}
\caption{$n=l=k=M=q=a=b=\eta_c=1$, $\alpha=0.5$, $\omega=0.5$}
\end{figure}
\begin{figure}
\centering
\includegraphics[width=3.0in,height=2.0in]{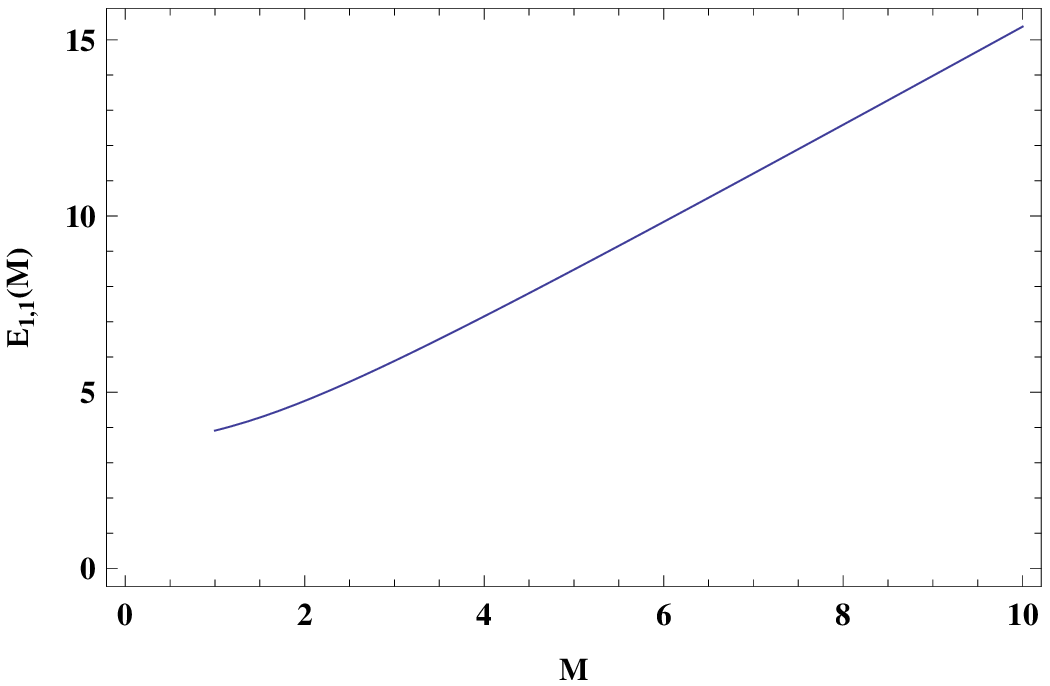}
\caption{$n=l=k=q=a=b=\eta_c=\eta_L=1$, $\alpha=0.5$, $\omega=0.5$}
\end{figure}
\begin{figure}
\centering
\includegraphics[width=3.0in,height=2.0in]{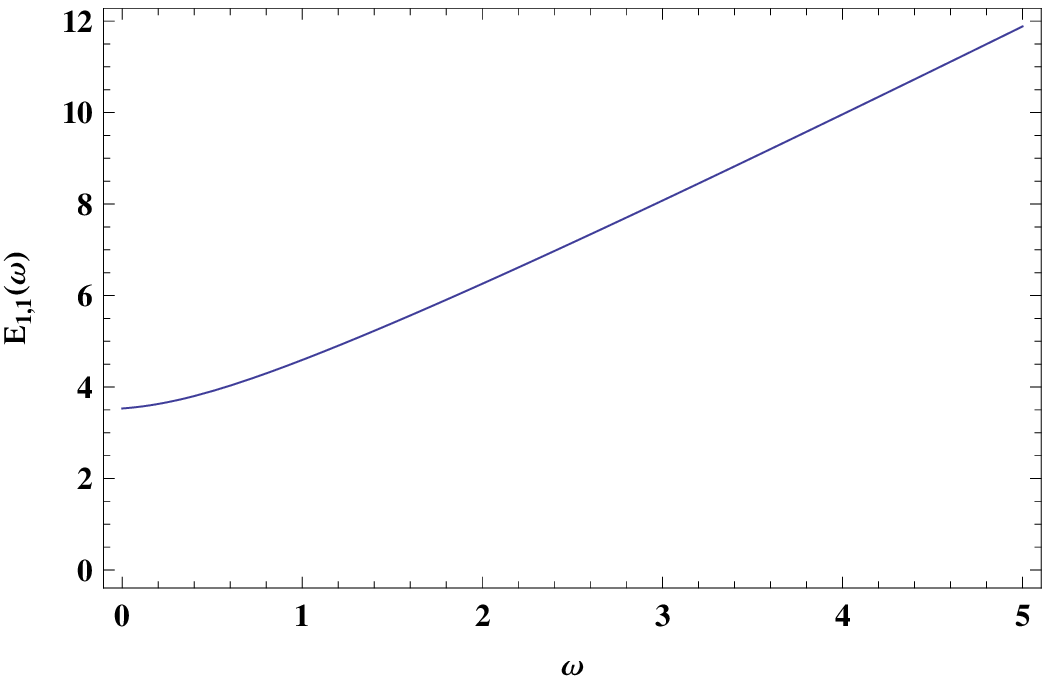}
\caption{$n=l=k=q=a=b=\eta_c=\eta_L=M=1$, $\alpha=0.5$}
\end{figure}
\begin{figure}
\centering
\includegraphics[width=3.0in,height=2.0in]{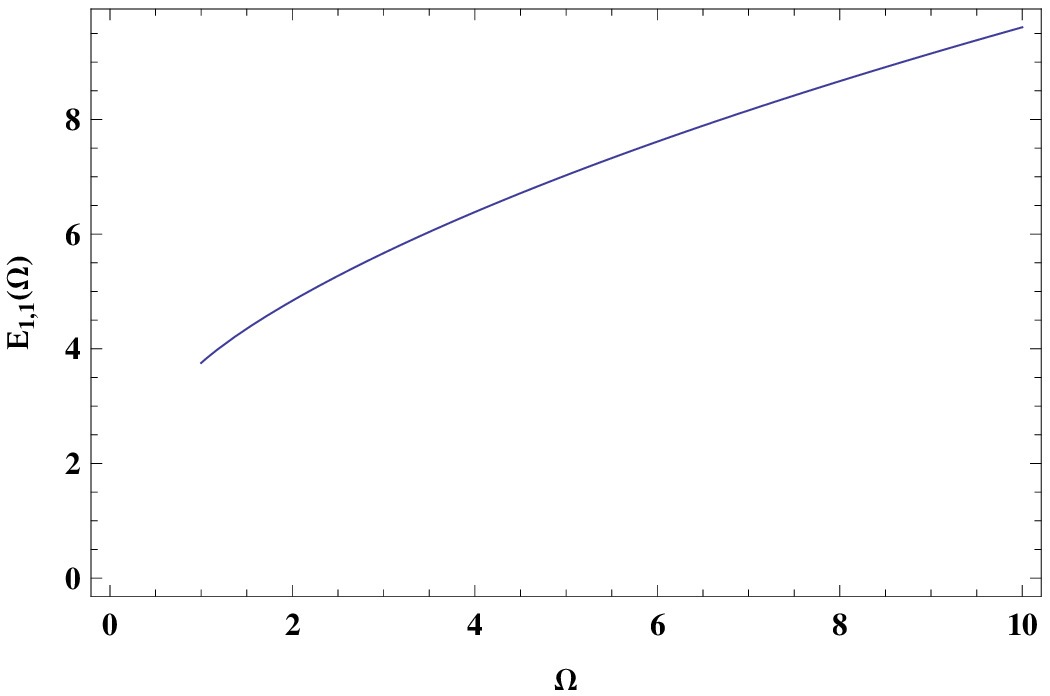}
\caption{$n=l=k=q=a=b=\eta_c=\eta_L=M=1$, $\alpha=0.5$, $\omega=0.5$}
\end{figure}
\begin{figure}
\centering
\includegraphics[width=3.0in,height=2.0in]{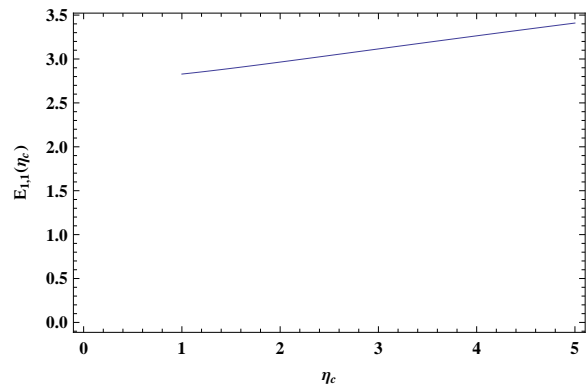}
\caption{$n=l=k=q=a=b=M=1$, $\alpha=0.5$, $\omega=0.5$, $\Phi=\pi$}
\end{figure}
\begin{figure}
\centering
\includegraphics[width=3.0in,height=2.0in]{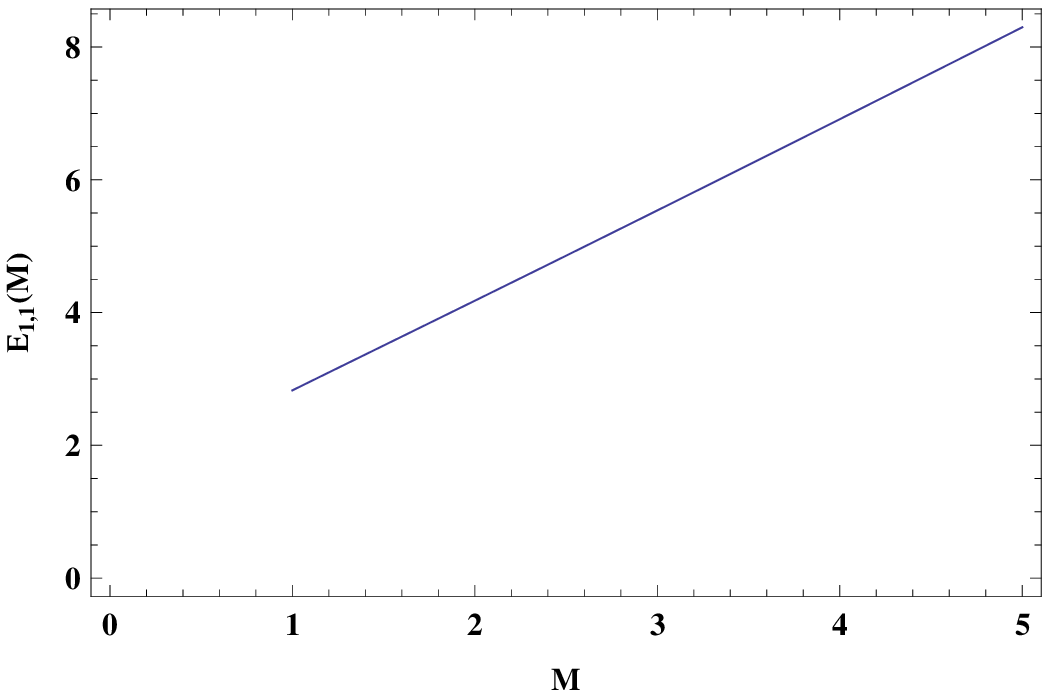}
\caption{$n=l=k=q=a=b=\eta_c=1$, $\alpha=0.5$, $\omega=0.5$, $\Phi=\pi$}
\end{figure}
\begin{figure}
\centering
\includegraphics[width=3.0in,height=2.0in]{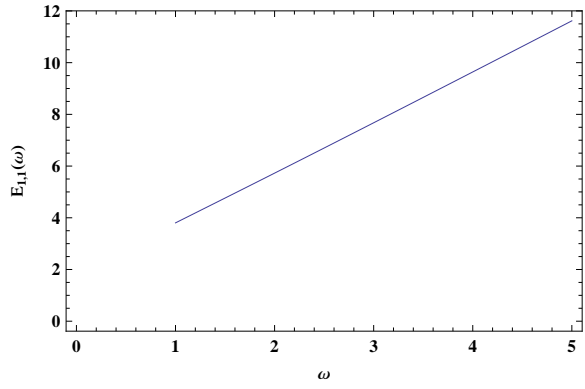}
\caption{$n=l=k=q=a=b=\eta_c=M=1$, $\alpha=0.5$, $\Phi=\pi$}
\end{figure}
\begin{figure}
\centering
\includegraphics[width=3.0in,height=2.0in]{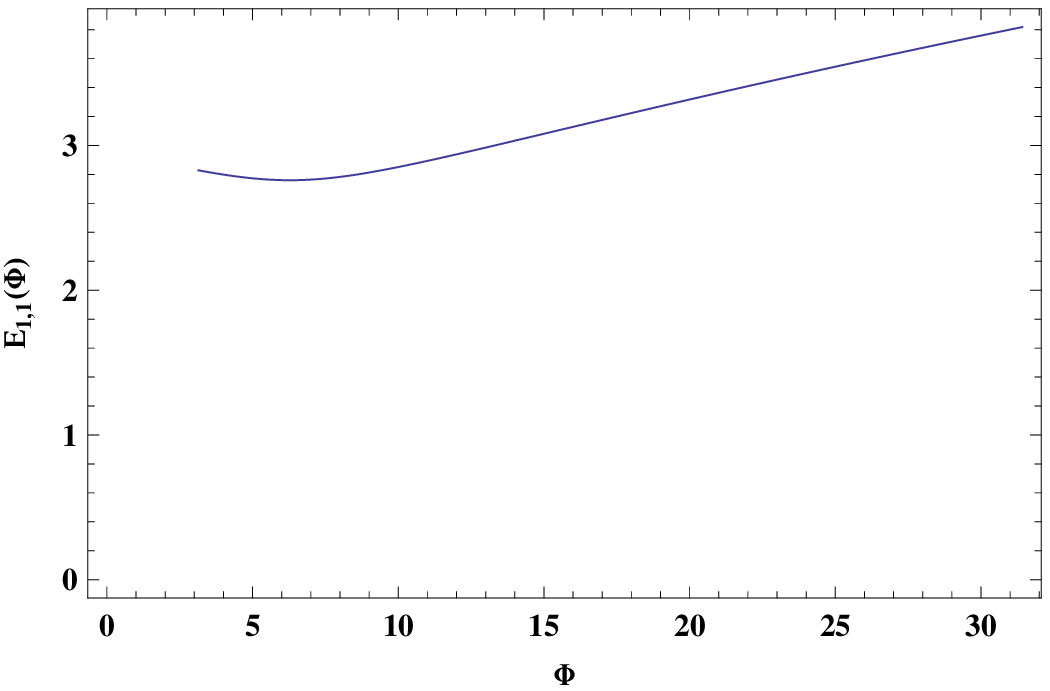}
\caption{$n=l=k=q=a=b=\eta_c=M=1$, $\alpha=0.5$, $\omega=0.5$}
\end{figure}


\end{document}